 \documentclass[preprint,12pt]{elsarticle}

\usepackage{amssymb}

\usepackage{dblfloatfix}
\usepackage{amsmath,amssymb,amsfonts}
\usepackage{algorithmic}
\usepackage[graphicx]{realboxes}
\usepackage{subfigure}
\usepackage[export]{adjustbox}
\usepackage{textcomp}
\usepackage{xcolor}
\usepackage{booktabs}
\usepackage{multirow}
\usepackage{vcell}
\usepackage{comment}
\usepackage[hidelinks]{hyperref}

\journal{Graphical Models}

\begin{document}

\begin{frontmatter}



\title{\red{Evaluating and Comparing Crowd Simulations: Perspectives from a Crowd Authoring Tool}}

\author{Gabriel Fonseca Silva}
\author{Paulo Ricardo Knob}
\author{Rubens Halbig Montanha}
\author{Soraia Raupp Musse}
\affiliation[**]{organization={Virtual Humans Laboratory (VHLab), School of Tecnology},
            addressline={\\Pontifical Catholic University of Rio Grande do Sul, \\Av. Ipiranga, 6681}, 
            city={Porto Alegre},
            postcode={90619-900}, 
            state={RS},
            country={Brazil}}

\newcommand\red[1]{{#1}}
\newcommand\redto[1]{{\color{red}#1}}

\begin{abstract}

\red{Crowd simulation is a research area widely used in diverse fields, including gaming and security, assessing virtual agent movements through metrics like time to reach their goals, speed, trajectories, and densities.}
This is relevant for security applications, for instance, as different crowd configurations can determine the time people spend in environments trying to evacuate them. 
In this work, we extend WebCrowds, an authoring tool for crowd simulation, to allow users to build scenarios and evaluate them through a set of metrics. The aim is to provide a quantitative metric that can, based on simulation data, select the best crowd configuration in a certain environment. 
\red{We conduct experiments to validate our proposed metric in multiple crowd simulation scenarios and perform a comparison with another metric found in the literature.} 
The results show that \red{experts in the domain of crowd scenarios} agree with our \red{proposed} quantitative metric.
\end{abstract}




\end{frontmatter}


\newcommand\red[1]{{#1}}
\newcommand\redto[1]{{\color{red}#1}}

\section{Introduction}
\label{sec:introduction}

Since the pioneering work of Thalmann and Musse~\cite{musse1997model} on the domain of crowd simulation, many other methods have been proposed in both microscopic~\cite{pelechano2007controlling} and macroscopic~\cite{hughes2002continuum} points of view and, more recently, combining them both~\cite{silva2019bioclouds}. Ranging from the simulation of a crowd of people or robots~\cite{de2012simulating} to the modeling of personality for simulated virtual agents~\cite{knob2018simulating,durupinar2015psychological}, crowd simulation has a wide range of uses in different fields.

Regarding the application in the safety area, crowd simulation models can be used to simulate people's behavior in different spaces, such as sports arenas, 
nightclubs, and subway stations. In these scenarios, an important aspect concerns the evacuation planning of such environments~\cite{murphy2013evacSim, zhou2022evacuationSubway}. 
Indeed, crowd simulations can provide valuable data to help the decision-making of safety professionals when evaluating evacuation plans. This data can include a variety of metrics, such as the time required for everyone to leave the environment, their average speeds during the evacuation, and information about their traveled distances. In addition, simulations should include visual information, such as the trajectories of agents in the crowd and mapping areas with higher densities.

Despite its importance, identifying optimal evacuation plans is not a trivial task\red{, as} people's behavior in egress scenarios, both in real and virtual environments, depends on many conditions such as building layout and the placement of pillars and walls~\cite{berseth2015environmentOptimization,haworth2016pillarPlacement}.
In the present work, we propose using WebCrowds~\cite{silva2022webcrowds} as a simulation tool to generate crowd metrics for evaluating and comparing different scenarios~\cite{wolinski2014estimationEvaluation,daniel2021trajectoryQualityMetric}.
Our method is compared with a work in the literature, proposed by Cassol et al.~\cite{cassol2017evaluating}, which goal is to provide an evaluation equation that quantitatively evaluates an evacuation plan.
\red{In their work, the authors consider relevant aspects of the crowd, such as total evacuation time, average evacuation time, agent speed, and local agent density, to evaluate quantitatively a certain crowd configuration. In their case and in our case as well, a crowd configuration states for the distribution of agents in the environment and the goals and obstacles present in the environment.}

This work extends the authoring tool WebCrowds~\cite{silva2022webcrowds}
to allow users to build, evaluate, and compare different configurations of crowd scenarios. 
WebCrowds is a user-friendly web platform where users can define environments and various crowd configurations and scenarios. Therefore, our platform runs the specified simulations and generates simulation data that can be used to evaluate the crowd efficiency and also be used to compare various crowd alternatives to select the best one. As far as we know, this is a unique platform with all such characteristics.
\red{To evaluate the evacuation of the environments, we also propose a new quantitative evaluation metric ($\phi$) based on crowd features such as simulation time and average speeds. Furthermore, our method is evaluated by a group of experts in crowd scenarios (CS) 
who qualitatively assess our results.}
In short, the main contributions of this work, when compared with WebCrowds~\cite{silva2022webcrowds}, are as follows:

\begin{itemize}
    \item \textbf{Inclusion of new metrics}: In addition to the simulation results already present in WebCrowds (i.e., a simulation time metric, density map, and trajectories map), we included five quantitative metrics: average time, average speed, average distance walked, average local density, and an evaluation metric. 
    
    \item \textbf{Multiple Configurations}: WebCrowds, in its original version, allowed the user to build only one configuration and scenario to simulate. In this work, \red{we extended the original model and improved the user interface, allowing users to build up to four alternative configurations of each crowd scenario. Here, we considered the crowd scenario as one specific configuration where the environment has a certain surface area and a particular number of goals and agents.} The alternative configurations represent versions of such crowd scenarios where some features can be changed, e.g., the initial location of agents and the positions of obstacles. Having this data, WebCrowds can simulate all defined configurations and compare them to select the best one.
    
    \item \textbf{Evaluation and Comparison}: In the original version, WebCrowds only \red{generates} visual information about a certain crowd simulation. In the present version, with the possibility to assemble up to four different configurations alongside the new evaluation metric included, it is possible to evaluate and compare different crowd scenarios to find the best configuration.
\end{itemize}

This paper is structured as follows. Section~\ref{sec:related} presents some work that aim to build crowd simulation tools and evaluate crowd behavior. Section~\ref{sec:model} presents the new features included in the original model of WebCrowds~\cite{silva2022webcrowds}. Section~\ref{sec:results} details the experiment conducted and the results achieved. Finally, Section~\ref{sec:finalConsiderations} presents our final considerations.

\section{Related Work}
\label{sec:related}

This section discusses some work on authoring tools and techniques that aim to evaluate crowd behaviors.

\subsection{Authoring Tools}
\label{sec:relatedAuthoring}

Although some methods in the literature aim to author crowd scenarios, we did not find any that specifically describe tools to simulate multiple crowd scenarios and simultaneously provide an evaluation and comparison among them. For instance, Ulicny et al.~\cite{DBLP:conf/siggraph/UlicnyCT05} presented a method of drawing a crowd simulation where the user could use different types of brushes to create the crowds and change their appearance, behavior, and direction to create a scene to simulate. The authors propose an exciting tool for authoring crowds; however, there is no evaluation of the generated scenario.
Chen et al. \cite{DBLP:journals/jvca/ChenWL20} presented a way to compose crowd scenarios using natural language processing techniques, where different words and expressions represent the attributes and behavior of each agent. Some of these expressions may contain emotions, such as "walk happily" and "walk sadly". 
This model was also employed by Liu et al.~\cite{DBLP:journals/jvca/LiuWC20}, where keywords such as quantity, character type, behavioral verb, location, and object are considered. The data structure of the crowd scene is divided into three components: crowd, behavior, and destination. Each of these components has its keywords. 

Colas et al.~\cite{sketchingFields2022} presented a model for authoring crowd simulations using sketches. The user draws curves in the interface, which compose an interaction field grid, influencing the steering behavior of agents. 
Kim and Sung~\cite{skecthingAR2022} presented a model using sketching in combination with augmented reality, where users can use dry-erase markers to compose scenes. The authors use computer vision techniques to identify shapes and colors in the sketches to define which object type (e.g., goals, paths, and trees) should be placed in that position. 
\red{Additionally, Lemonari et al.~\cite{lemonari2022authoring} presented a survey of many authoring tools, where they provide a review of the most relevant methods in each area, discussing the available authoring tools while identifying the trends of early and recent work, as well as suggesting promising directions for future research.}

The mentioned methods focus on crowd animation, where users may define the behaviors of virtual characters. The following section presents related works on crowd evaluation.

\subsection{Crowd Evaluation}
\label{sec:relatedCrowdEval}

Cassol and collaborators~\cite{cassol2017evaluating} proposed an evacuation metric that can identify optimal evacuation plans given a set of different scenarios. The proposed metric was built considering relevant aspects of the crowd behavior in evacuation scenarios, such as total evacuation time, average evacuation time, agent speed, and local agent density. To validate their model, the authors used the model of a nightclub, including data from a real-life evacuation, to compare the results.

Focusing on the layout of a building, the work of Schaumann et al.~\cite{schaumann2019simulating} explores the impact that a building design can make on people inside this building. To do so, the authors developed a narrative-based pre-occupancy evaluation platform that supports the simulation of human behavior and the impact of the environment on their behavior. Focused on the environment of a hospital, the platform was applied to conduct a comparative evaluation of two different architectural designs for an outpatient ophthalmology clinic. The results demonstrate the potential of narrative-based modeling and how design decisions may impact future building operations.

The concept proposed by Usman et al.~\cite{DBLP:conf/mig/UsmanLMZFK20} allows the comparison of an environment with different agent behaviors in a social distance situation. The authors used a metric to calculate a Social Distancing Index (SDI), an environmental performance measure. Each agent has a 1-meter radius circle called social space, and when an agent circle overlaps another agent's social space, we have a social distance violation. Besides the SDI, some of the metrics presented by the authors are total violations, cell violations, unique violations, and average cell violations.

In addition, Usman et al.~\cite{DBLP:journals/jvca/UsmanHFK21} suggest different ways of presenting simulation feedback. Using Software-as-a-service (SaaS) and running an authoring tool, the authors define some environment specifications, such as an Industry Foundation Classes (IFC) file and a Level of Service (LoS), which defines the crowd density. The software presents dynamic crowd analysis and visualization results with this input data, which show a path and bottleneck analysis. The path analysis allows the user to see the crowd movement trajectories in a static view or in a dynamic way, where he/she can see the routes as an animation or in a specific timestep. Similar to the path analysis, the bottleneck analysis is a color-coded heatmap showing the density contours, where areas in red represent high density (potential bottlenecks) and areas in blue represent less density. They also provide quantitative information, such as exit flow, distance walked, and simulation time. Yet, cultural behaviors can also be considered in simulation as proposed in the work by Favaretto~\cite{Fav2017}.

\red{The literature's work presented in Section~\ref{sec:relatedAuthoring} 
is concentrated on crowd authoring and animation, where new user interaction methods, like natural language and augmented reality, are explored to define behaviors and parameters in a crowd.} 
Also, the literature presented in this section 
defines ways of evaluating crowd evacuation or behavior scenarios. The different metrics presented can allow users to evaluate the evacuation plan or social distance violations. The literature also presents tools that return visualization results, presenting visual analysis. Unlike them, WebCrowds provides an authoring tool to model crowd scenarios, including environment modeling
and population motion, allowing users to evaluate virtual agents’ trajectories and densities through visual information and, also, providing quantitative data \red{and} comparison with alternative scenarios.


\section{Proposed Model}
\label{sec:model}

In this section, we present our proposed model for authoring and evaluating multiple crowd simulation scenarios.

\subsection{System Overview}
\label{sec:system_architecture}

\begin{figure*}[!htb]
    \centering
    \includegraphics[width=0.98\textwidth]{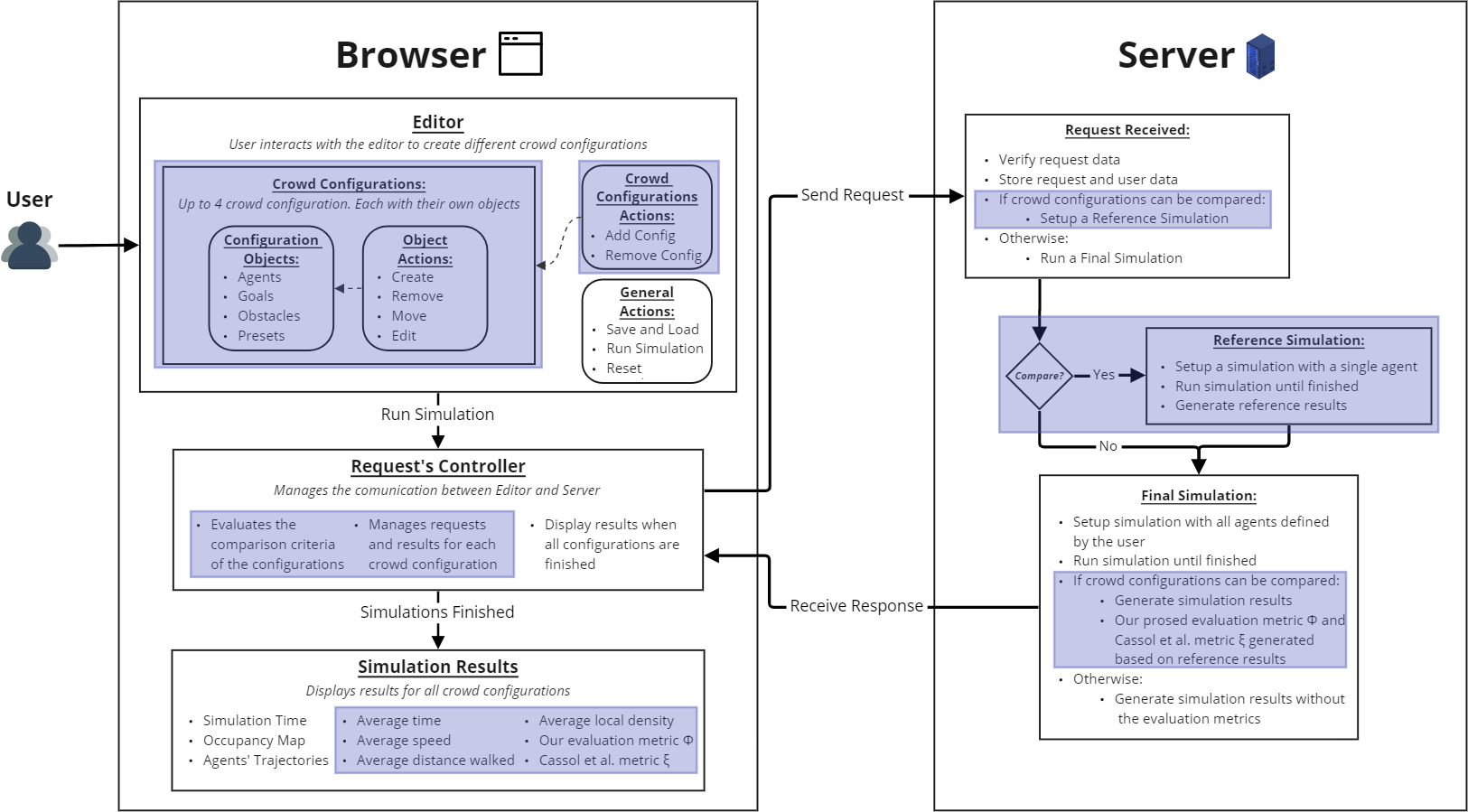}
    \caption{
    The pipeline of WebCrowds. Users can use the Editor through a web browser to create simulation scenarios. Requests for a given simulation are sent to the Server, responsible for executing the simulation, generating results, and sending them back to be shown in the Browser. Elements highlighted in purple are new features in our proposed model, which include the possibility of creating multiple environments to be evaluated and compared through metrics.}
    \label{fig:overview}
\end{figure*}

Figure~\ref{fig:overview} presents an overview of the proposed model, with new features that focus on evaluating and comparing multiple crowd scenarios highlighted in purple. Our tool is available online\footnote{Available at: \url{ https://vhlab.com.br/projects/webcrowds_comparable/}.}. 

The pipeline of execution follows the structure of the original work~\cite{silva2022webcrowds}. 
First, users can interact with the Editor to build simulation scenarios with different crowd configurations. Each Crowd Configuration has its objects, which include the spawn areas of virtual agents, their goals (i.e., exits), obstacles, and preset (pre-made sets of obstacles). The user can modify the objects of each configuration independently using the Object Actions, which includes options to create, remove, move, and edit them. Crowd Configuration Actions allow users to add a new crowd configuration (up to 4 in our current implementation) or remove an existing one. The General Actions allow them to save the current crowd configurations in a file for future use and run the simulations using all configurations.

When the user runs the simulations, the Request's Controller gathers information from all crowd configurations and evaluates if they can be compared. The comparison criteria are described in Section~\ref{sec:evaluationConfig}. Then, the controller individually sends information about each configuration to be simulated in the Server, alongside a flag telling if the different crowd configurations can be compared. 
When the Server receives a simulation request, it verifies and stores the data. If the crowd configurations can be compared, the Server will set up and run a Reference Simulation for each configuration, generating reference results. The Reference Simulation and its results are described in Section~\ref{sec:metrics}. Then, a Final Simulation is performed, generating results that include a set of simulation metrics and visualizations. If the crowd configurations can be compared, the results will also include our proposed evaluation metric $\phi$, calculated using data from the reference simulation. All simulation metrics, including the evaluation metric $\phi$, are described in Section~\ref{sec:metrics}. Once the simulations of all crowd configurations are finished, the Request`s Controller sends the Simulation Results data to be displayed on the browser.

\subsection{Simulation Metrics}
\label{sec:metrics}

As defined in the original work~\cite{silva2022webcrowds}, WebCrowds is an authoring tool for crowd simulation that delivers a set of simulation results back to the user so they can evaluate whether the simulated scenario is satisfactory or not.
The set of simulation results was compounded by a total simulation time metric ($t_g$), 
which represents the time required for all agents to achieve their goals, and two visualizations: one with a density map (now named ``occupancy map''), representing the number of agents that walked through a particular area of the space, and a trajectories map, which plots the positions and goals of all agents during the simulation. Figure~\ref{fig:simulation_scenarios_part1} and Figure~\ref{fig:simulation_scenarios_part2} presents examples of the occupancy map and agents' trajectories.
For more details about these simulation results, please consult the original work~\cite{silva2022webcrowds}.

To compare different environment crowd scenarios, we included additional metrics to be presented to the users at the end of the simulations.
We based the additional metrics on the work of Cassol et al.~\cite{cassol2017evaluating}, where the authors proposed an evaluation metric for evacuation plans. This metric is based on important aspects of the crowd simulations, such as the time taken for agents to evacuate, agents' speed, and the density of the agents. 

In the present work, in addition to the original metric of WebCrowds, we included the following ones:

\begin{itemize}
    \item \textbf{Average time ($\bar{t}$)}: Besides the total time of simulation $t_g$ (already included in the original work), we also included the average time $\bar{t}$ that the agents took to reach their respective goals, defined by Equation~\ref{eq:averageTime}:    
    \begin{equation}
    \label{eq:averageTime}
        \bar{t} = \frac{\sum_{k=1} ^{A} t_k}{A},
    \end{equation}
    where $A$ is the total quantity of simulated agents, and $t_k$ is the time agent $k$ takes to reach its goal.
    
    \item \textbf{Average density ($\bar{d}$)}: In addition to the visual information provided by the occupancy map, we also included the average density $\bar{d}$, defined by the following equations:    
    \begin{equation}
    \label{eq:averageDensity}
        C = \sum_{i=1}^{Cells} X_i, \\
    \end{equation}
    
    \begin{equation}
    \label{eq:averageDensity2}
        X_i = \begin{cases}
        1,  & \text{if $Ag_i>0$} \\
        0, & \text{otherwise} 
        \end{cases}
    \end{equation}
    
    \begin{equation}
    \label{eq:averageDensity3}
        \bar{d} = \frac{A}{C},
    \end{equation}
    \red{where $Cells$ \red{stands} for the number of 1x1 sqm cells (i.e., square meters) in the environment. Then, we count how many cells $C$ are occupied by more than 0 agents ($Ag$) in Equation~\ref{eq:averageDensity}. Finally, we compute $\bar{d}$ as the number of agents in the simulation ($A$) divided by the number of occupied cells ($C$) (Equation~\ref{eq:averageDensity3})}.
    
    \item \textbf{Average speed ($\bar{s}$)}: We computed the average speed $\bar{s}$ of all the agents present in the simulation, defined by the Equation~\ref{eq:averageSpeed}:    
    \begin{equation}
    \label{eq:averageSpeed}
        \bar{s} = \frac{\sum_{k=1} ^{A} s_k}{A},
    \end{equation}
    where $s_k$ is the average speed of agent $k$.
    
    \item \textbf{Average distance walked ($\bar{w}$)}: In the present work, we included the average distance walked $\bar{w}$ by all the agents in the simulation, defined by the  Equation~\ref{eq:averageDistance}:    
    \begin{equation}
    \label{eq:averageDistance}
        \bar{w} = \frac{\sum_{k=1} ^{A} w_k}{A},
    \end{equation}
    where $w_k$ is the total distance walked by agent $k$.
\end{itemize}

In addition to the new metrics mentioned above, we also included a quantitative evaluation metric to evaluate the simulation as a whole. To do so, we also refer to the work of Cassol et al.~\cite{cassol2017evaluating}, where a similar metric was proposed. 
Cassol's metric is composed of four aspects of the crowd: the total simulation time ($t_g$), the average time of simulation ($\bar{t}$), average density ($\bar{d}$), and average speed ($\bar{s}$). While $t_g$ was already present in WebCrowds~\cite{silva2022webcrowds}, $\bar{t}$, $\bar{d}$ and $\bar{s}$ are included in this work and calculated as presented in Equations~\ref{eq:averageTime},~\ref{eq:averageDensity3} and~\ref{eq:averageSpeed}, respectively. 
Additionally, we propose in the present work that the distance walked by the agents during the simulation is also a relevant feature of the crowd. The reason is that people usually try to reach their respective goals following the \red{least} effort path~\cite{still2000crowd, fu2014exitSelection}, which includes walking \red{as little as possible} to achieve the goals. Thus, we included the average distance walked $\bar{w}$, presented in Equation~\ref{eq:averageDistance}. 

Cassol et al.~\cite{cassol2017evaluating} considered that the complexity of the environment affects the evaluation metric. To do so, an additional simulation is run with a single agent, called the reference agent (highlighted as the ``Reference Simulation'' in Figure~\ref{fig:overview}). \red{The data generated by this simulation is used to normalize the evaluation metrics by adjusting the complexity of the scene present as a function of the environment and not by the presence of other agents. There are two exceptions: the first 
is the normalization of the density value, since the density of one agent (reference agent) is always constant and does not depend on the environment's configuration. The second is the normalization of the distance, proposed in the present work. The reason is that the distance depends on the agents' locations at the beginning of the simulation, and we wanted to have a normalization only dependent on the environment. Thus, the normalized values for the simulation metrics are calculated as follows:}

\begin{equation}
\label{eq:normalTotalTime}
    t^{'}_g = \frac{t_g}{t_{ar}},
\end{equation}

\begin{equation}
\label{eq:normalAverageTime}
    \bar{t^{'}} = \frac{\bar{t}}{t_{ar}},
\end{equation}

\begin{equation}
\label{eq:normalAverageSpeed}
    \bar{s^{'}} = \exp(\frac{s_{ar}}{\bar{s}}),
\end{equation}

\begin{equation}
\label{eq:normalDistanceWalked}
    \bar{w^{'}} = \frac{\bar{w}}{a_{s}},
\end{equation}
where $t_{ar}$ is the simulation time for the reference agent, $s_{ar}$ is the average speed of the reference agent
and $a_{s}$ is the \red{hypotenuse of the simulation's environment, represented as a quadrilateral.} 
\red{For Equation~\ref{eq:normalAverageSpeed}, we used the $exp()$ function following the method adopted by Cassol et al.~\cite{cassol2017evaluating}, where the idea was to give less importance to the speed when compared to the other metrics. It was done so because it will generate a smaller value when the prime factor is computed in the harmonic mean (Equation~\ref{eq:evacuationMetric}).}
\red{For Equation~\ref{eq:normalDistanceWalked}, our aim was to use a value greater than any distance an agent could traverse in an environment with minimal obstacles (i.e., essentially, a simpler environment). In other words, our intention was to incorporate implicit information about the complexity of the environment while reducing dependence on the reference agent's traveled distance, which is imperative due to the differing maximum achievable distances for each scenario. Additionally, this metric is contingent on the complexity of the environment. We assessed that configurations with a smaller average distance (Equation~\ref{eq:averageDistance}) than the hypotenuse would indicate a less complex environment. In situations with simpler environments, we aimed to give greater significance to the distance disparity to evaluate the various crowd configurations. Thus, in configurations where agents cover less distance, the prime distance metric is smaller $<$ 1 (see Equation~\ref{eq:normalDistanceWalked}), consequently this factor $\frac{1}{\bar{w^{'}}}$ impacts more in the Harmonic equation (see Equation~\ref{eq:evacuationMetric}) to configurations where agents cover greater distances.
Finally, the evacuation metric $\phi$, proposed in the present work, is calculated as a harmonic mean, as follows:}

\begin{equation}
\label{eq:evacuationMetric}
    \phi = \frac{5}{\frac{1}{t^{'}_g} + \frac{1}{\bar{t^{'}}} + \frac{1}{\bar{d}} + \frac{1}{\bar{s^{'}}} + \frac{1}{\bar{w^{'}}}},
\end{equation}
where the lower the value of $\phi$, the better the scenario evaluation. 
\red{As commented before, we based our work on the metric proposed by Cassol et al.~\cite{cassol2017evaluating}; thus, that is where the variables density, global time, local time, and speed came from. However, the authors of that work did not have an interactive way and authoring tool to test their metric in many different environments. So, in our work, we decided to test the metric in our authoring tool by designing several different environments. Based on the empirical studies of such metric in our platform, we found out that the variable of distance walked ($\bar{w^{'}}$) was also relevant, so we proposed an extension to the metric. 
We did not change the four metrics proposed by Cassol et al.~\cite{cassol2017evaluating}, where all prime values are greater than 1, so values in the Equation~\ref{eq:evacuationMetric} 
have the same weight/importance in calculating the evacuation metric $\phi$. 
In the case of the new traveled distance metric, we proposed that it can have a prime value smaller than one ($\bar{w^{'}}<1$) if the environment is simple. \red{Since $\bar{w^{'}}$ is incorporated in the Harmonic equation (Equation~\ref{eq:evacuationMetric}), it should have a greater impact on comparisons of crowd configurations.}
In the cases where the environment is complex, it would have a prime value greater than one ($\bar{w^{'}}>1$), like other metrics.}


\subsection{Evaluation and Comparison of Multiple Configurations}
\label{sec:evaluationConfig}

\begin{figure*}[!htb]
  \centering
  \subfigure[fig:example_criteria_a][Configuration A]
  {\includegraphics[width=0.31\textwidth]{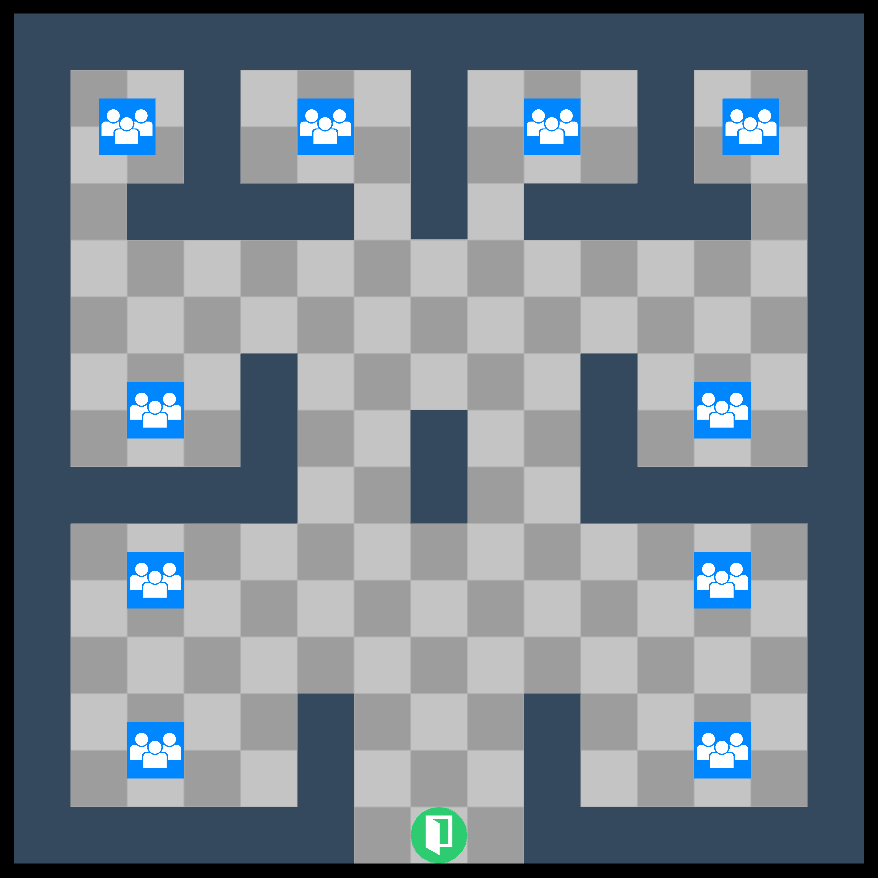}
  \label{fig:example_criteria_a}}
  \subfigure[fig:example_criteria_b][Configuration B]
  {\includegraphics[width=0.31\textwidth]{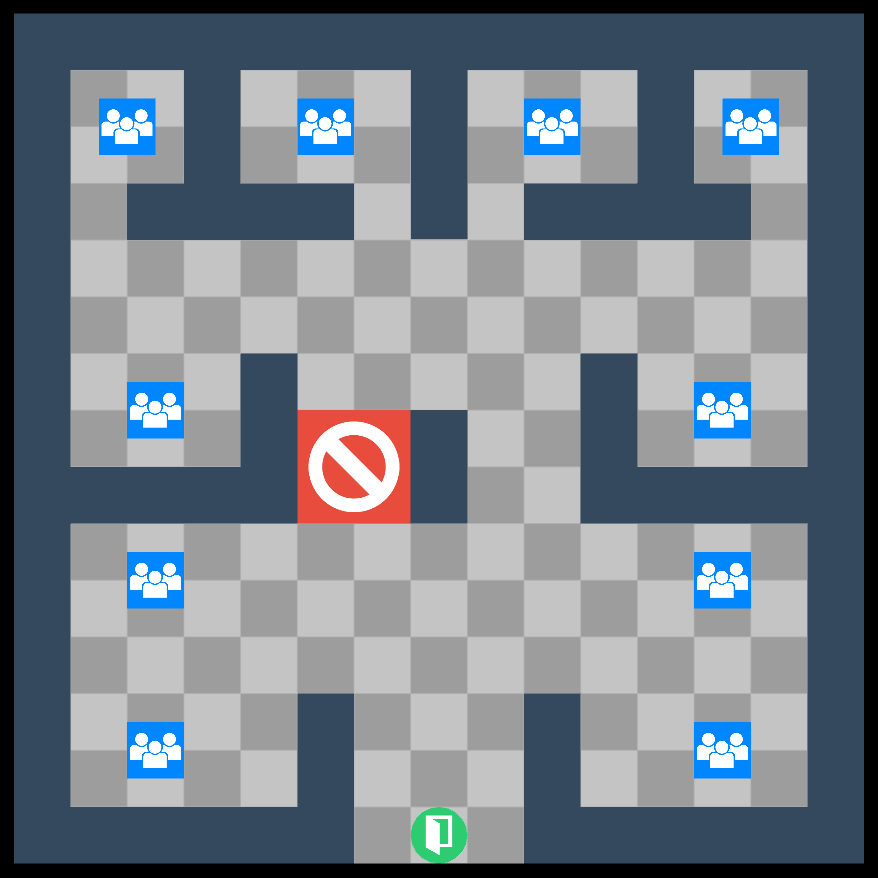}
  \label{fig:example_criteria_b}}
  \subfigure[fig:example_criteria_c][Configuration C]
  {\includegraphics[width=0.31\textwidth]{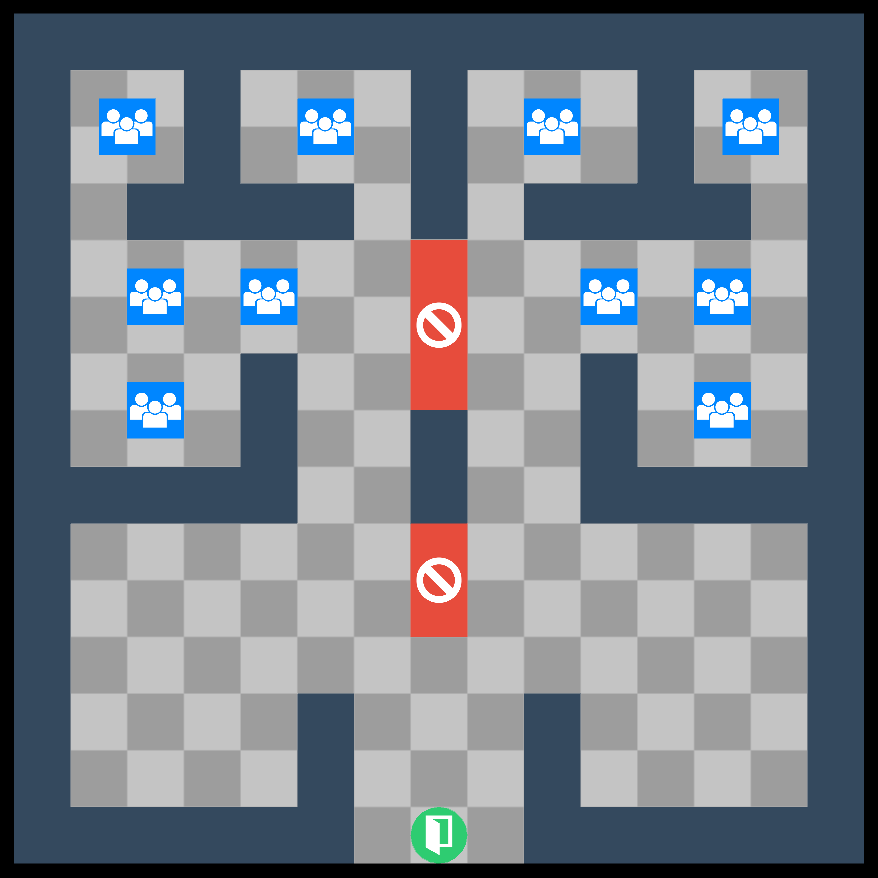}
  \label{fig:example_criteria_c}}
  \caption{Example of a crowd simulation scenario with three different configurations. Blue squares represent the agents' spawn areas, the green circle at the bottom represents the agents' goal, and obstacles are represented in red. 
  In this case, the configurations met the comparison criteria, as the total number of agents (although in different positions), the number of goals, and the environment geometries are the same. 
  }
 \label{fig:example_criteria}
\end{figure*}

To allow the comparison of different scenarios using the evaluation metric $\phi$, WebCrowds was extended to permit users to edit multiple crowd configurations as desired. To achieve that, a deep copy of the original environment is made, copying the position of spawn areas, goals, and obstacles into a new environment configuration. This process also copies additional information about these objects, such as the size and rotation of obstacles, the number of agents per spawn area, and their respective goals.
In our current implementation, the user interface is limited to a maximum of four different configurations; however, this is not a limitation of our model, as it could receive any number of structures to be simulated. Figure~\ref{fig:example_criteria} presents a \red{simulation scenario} with three different configurations.

As mentioned by Cassol and collaborators~\cite{cassol2017evaluating}, not all crowd scenarios and environments make sense to be compared. For instance, a building with 100 or 1000 people are supposed to have different evacuation data. The same happens with two different buildings, i.e., comparing evacuation data of two distinct spaces does not make sense.
So, in our case, we defined some requirements that must be achieved to provide crowd comparisons: 

\begin{itemize}
    \item \textbf{Same total number of agents.} The number and position of spawn areas can be modified as long as the total number of agents created in each configuration is the same. For example, a specific crowd configuration may contain one group of 10 agents, while another configuration may contain two groups of 5 agents.
    
    \item \textbf{Same number of goals (i.e., exits).} Similar to the number of agents requirement, the positions of each goal can be different in each configuration as long as the quantity of goals is the same.
    
    \item \textbf{Same surface area.} All configurations from a scene must contain the same environment geometry, i.e., the same rectangle~\footnote{The rectangle is the only shape accepted for the environment in the present prototype. The users can design more complex spaces using obstacles.} dimensions, 
    as it is used to normalize the average distance walked $a_{s}$. Still, the quantity, size, and position of obstacles can differ among the designed configurations.
\end{itemize}

When all criteria are met (see an example in Figure~\ref{fig:example_criteria}), the different configurations can be compared, and the evaluation metric $\phi$ will be presented as part of the simulation results. This requires the execution of a reference simulation for each configuration, using the reference agent data to calculate $\phi$. In our implementation, the agent furthest from its goal, in one of the spawn areas, is selected as the reference agent for each configuration. One can say that this choice can impact the results. However, we justify that once this decision is taken for all simulations, the impact is similar to all of them, keeping them comparable.

Still, in terms of usability, users can modify any configuration as desired, which could violate one or more criteria. In this case, i.e., any mentioned criteria not being met, the crowd scenarios are considered not comparable, 
so all 
metrics defined in Section~\ref{sec:metrics} are presented to users, except the evaluation metric $\phi$.

\section{Experimental Results}
\label{sec:results}

In this section, we present the results achieved by our experiments. Section~\ref{sec:metricValidation} presents the investigation conducted to validate the new evaluation metric $\phi$ proposed in this work. Section~\ref{sec:metricComparison} compares our proposed evaluation metric $\phi$ and the evaluation metric $\xi$ presented by Cassol et al.~\cite{cassol2017evaluating}. Section~\ref{sec:focusGroup} presents the experiment conducted with a focus group of \red{experts in crowd scenarios}. 

\subsection{Evaluation Metric $\phi$}
\label{sec:metricValidation}

\begin{figure*}[!htb]
  \centering
  \subfigure[Simulation Scenario 1. Each configuration contains 90 virtual agents in an environment of 30 x 30 meters.]
  {\includegraphics[width=0.48\textwidth]{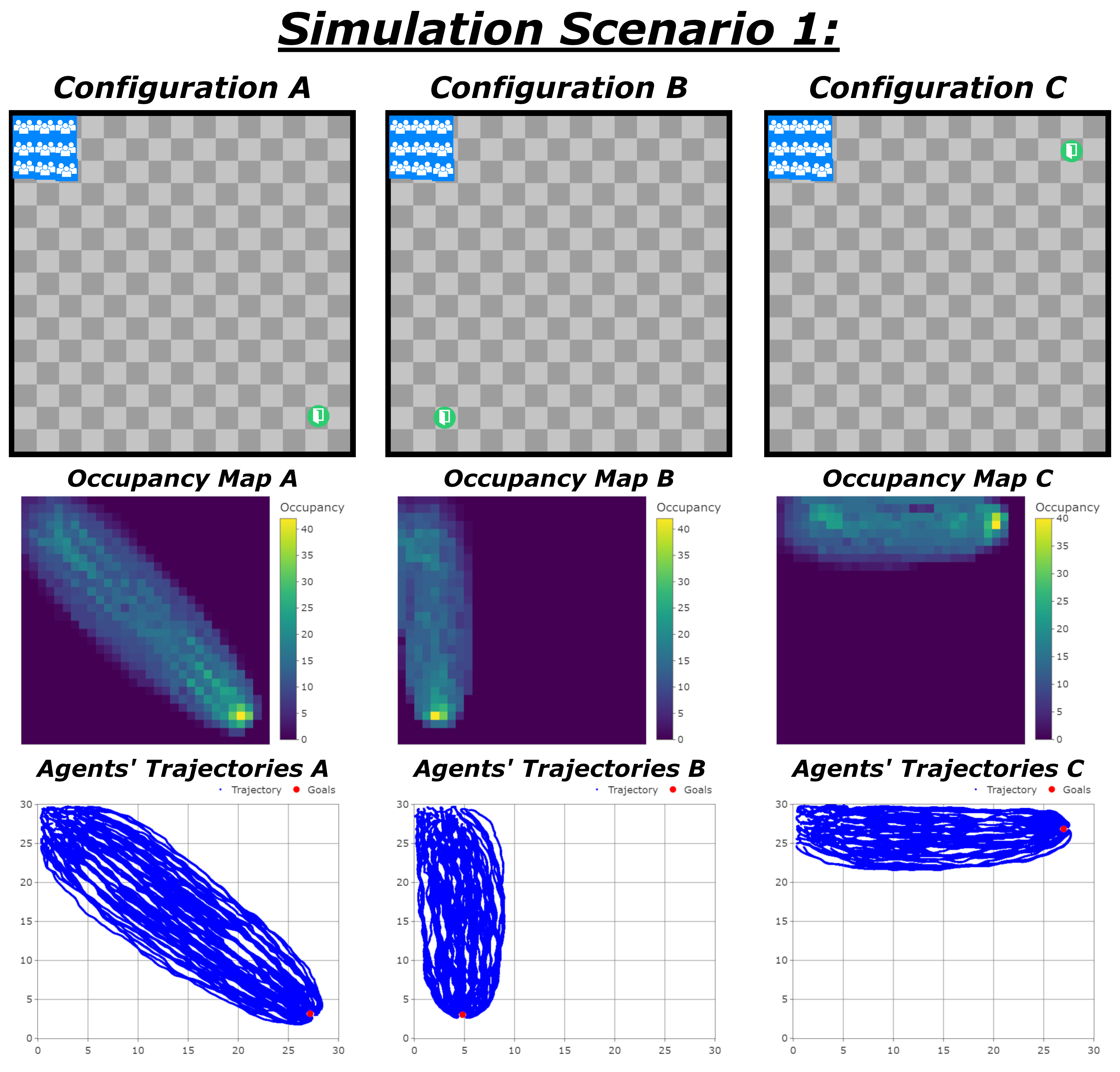}
  \label{fig:simulation_scenario_1}}
  \subfigure[Simulation Scenario 2. Each configuration contains 90 virtual agents in an environment of 30 x 30 meters.]
  {\includegraphics[width=0.48\textwidth]{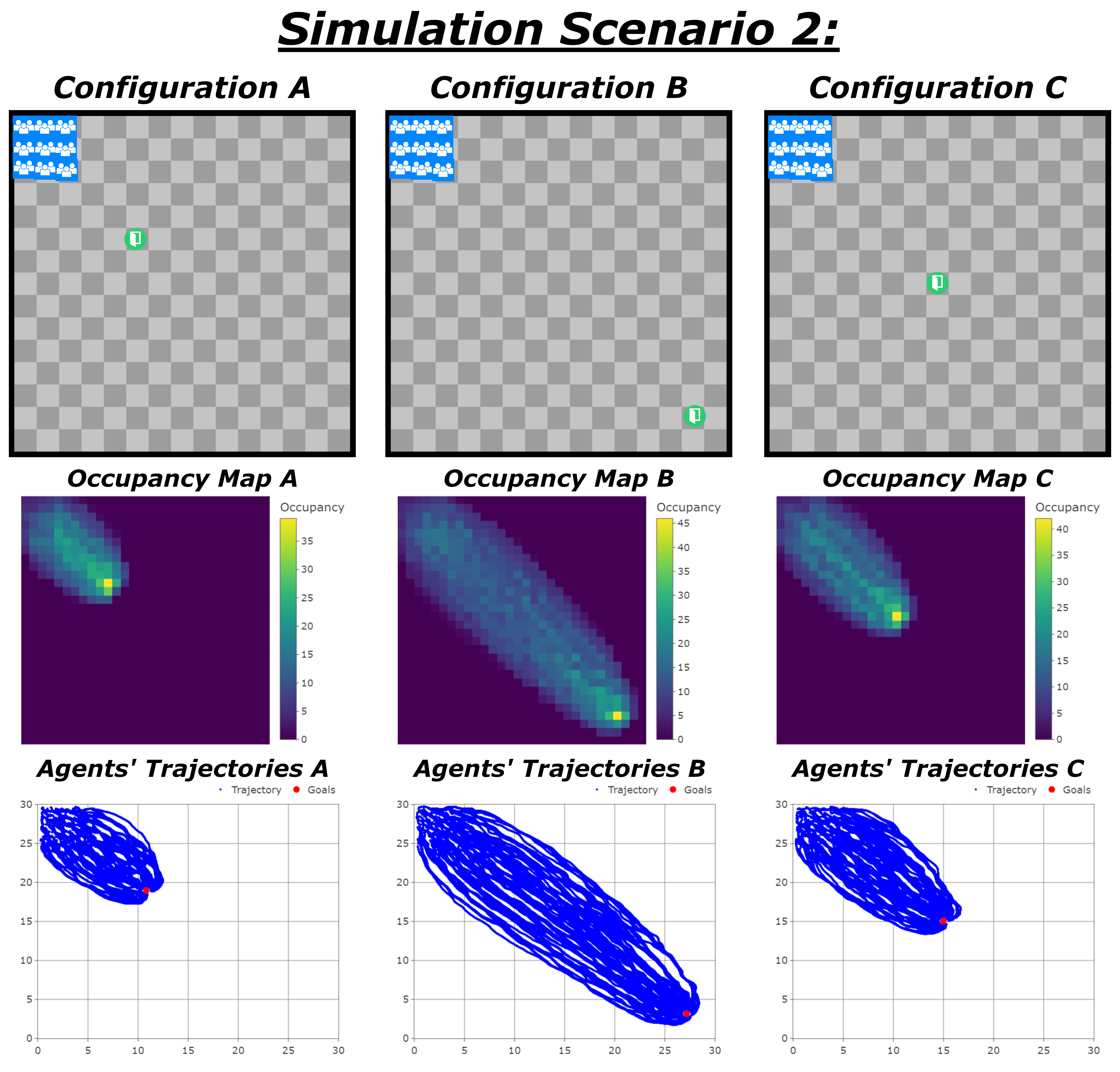}
  \label{fig:simulation_scenario_2}}
  \caption{Simulations Scenarios 1 and 2. The first line presents the different crowd configurations, with the initial position of agents (blue) and goals (green). These scenarios do not contain obstacles or walls. The second line presents the occupancy maps of each crowd configuration. The third line presents the plotted trajectories of virtual agents in each crowd configuration.}
 \label{fig:simulation_scenarios_part1}
\end{figure*}

\begin{figure*}[!htb]
  \centering
  \subfigure[Simulation Scenario 3. Each configuration contains 40 virtual agents in an environment of 30 x 30 meters.]
  {\includegraphics[width=0.48\textwidth]{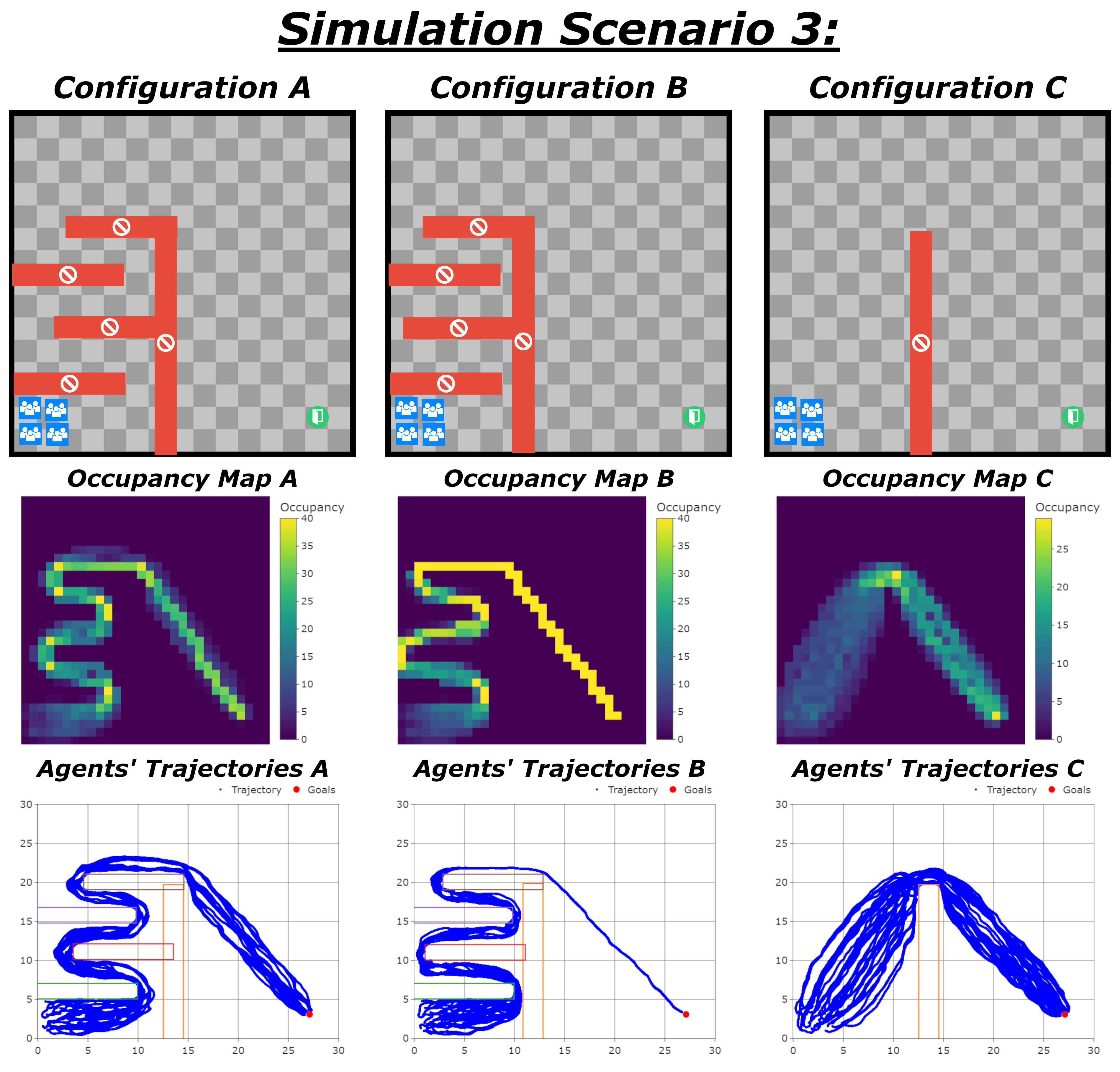}
  \label{fig:simulation_scenario_3}}
  \subfigure[Simulation Scenario 4. Each configuration contains 90 virtual agents in an environment of 30 x 30 meters.]
  {\includegraphics[width=0.48\textwidth]{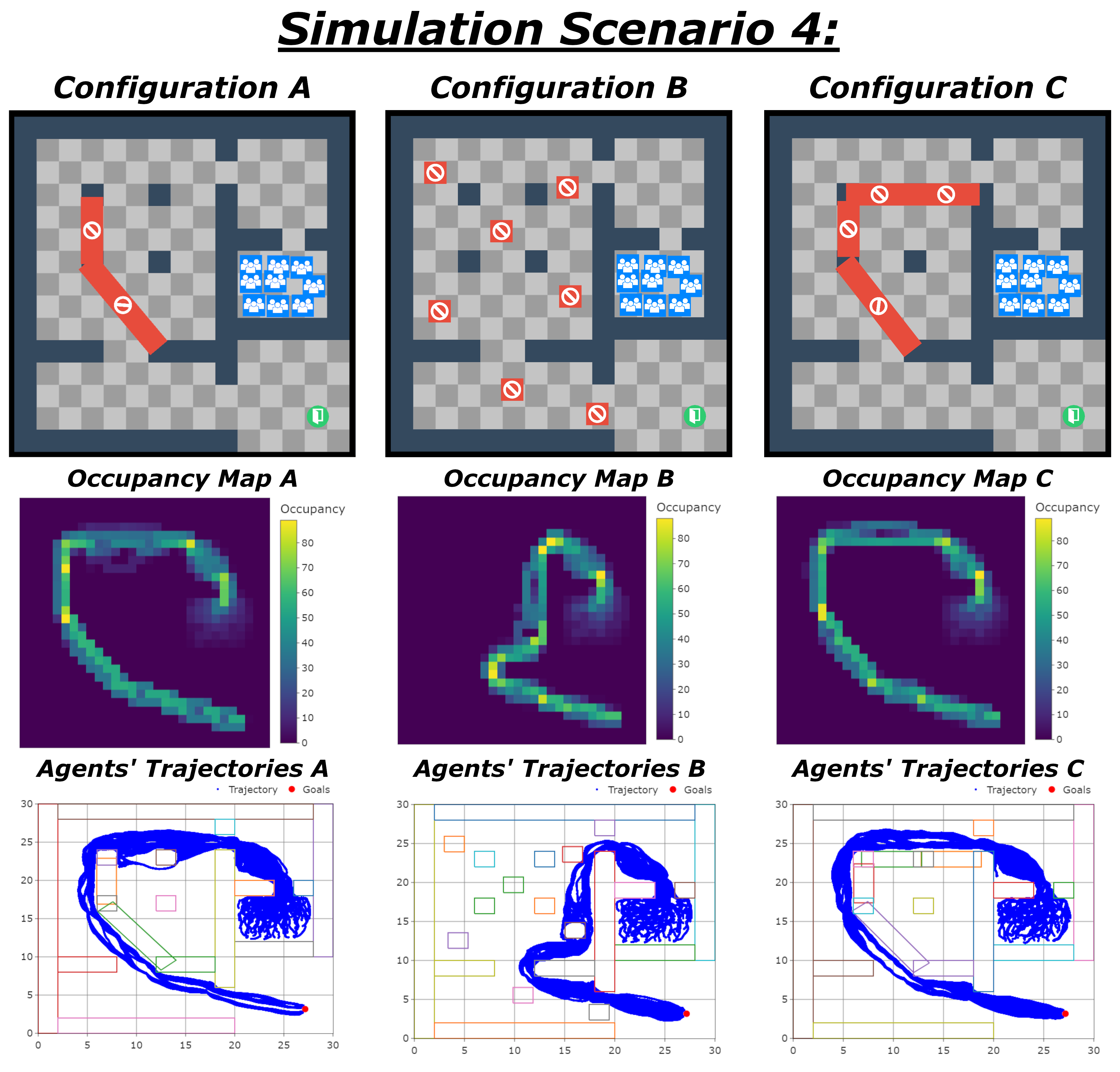}
  \label{fig:simulation_scenario_4}}
  \subfigure[Simulation Scenario 5. Each configuration contains 90 virtual agents in an environment of 60 x 15 meters.]
  {\includegraphics[width=0.60\textwidth]{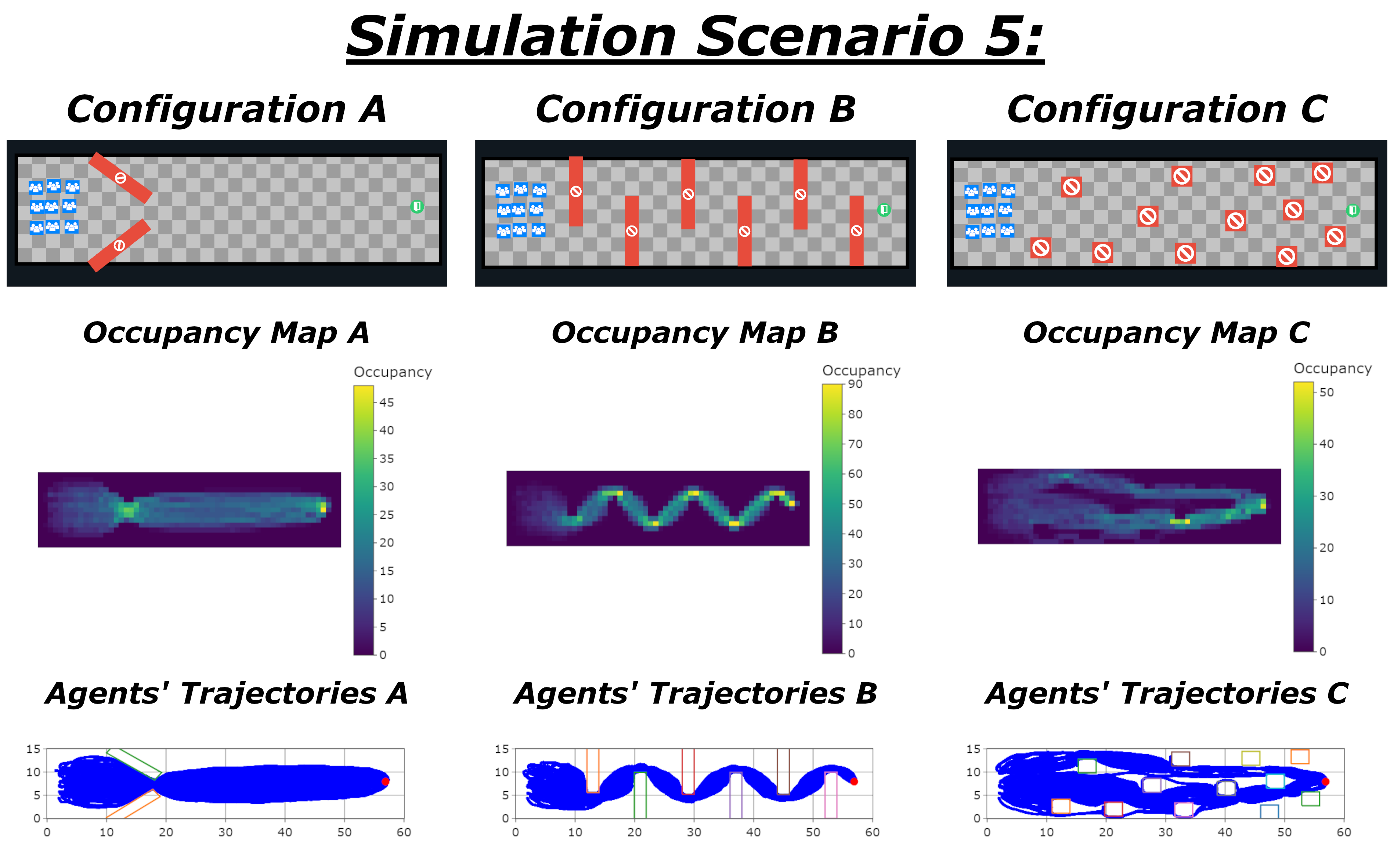}
  \label{fig:simulation_scenario_5}}
  \caption{Simulations Scenarios 3, 4, and 5. The first line presents the different crowd configurations, with the initial position of agents (blue), the goals (green), the obstacles (red), and the walls (dark grey). The second line presents the occupancy maps of each crowd configuration. The third line presents the plotted trajectories of virtual agents in each crowd configuration.}
 \label{fig:simulation_scenarios_part2}
\end{figure*}

First, we performed experiments to validate $\phi$. 
We modeled \red{five} crowd simulation scenarios (in Figures~\ref{fig:simulation_scenarios_part1} and~\ref{fig:simulation_scenarios_part2}), each containing three configurations that met the comparison criteria described in Section~\ref{sec:evaluationConfig}. 
\red{We performed 10 runs for every configuration (3 configurations for each of the 5 scenarios). Each run consists of a} Reference Simulation (considering only one agent) and a Final Simulation (considering the complete configuration scenario).
As shown in Figures~\ref{fig:simulation_scenarios_part1} and~\ref{fig:simulation_scenarios_part2}, 
for each scenario, in the first line, there is the initial position of agents (blue), the goals (green), the obstacles (red), and the walls (dark grey). The second line contains the occupancy maps generated as a function of crowd trajectories representing the achieved densities during the simulation. Finally, in the third line, we present the plotted trajectories.
Scenarios 1 and 2 explore the movement of agents in open spaces with exits at different distances and directions, while scenarios \red{3, 4, and 5} explore different obstacle layouts in more complex environments. \red{Scenarios 1 through 4 have an area of 30 x 30 meters, with Scenario 5 having an area of 60 x 15 meters}. Scenarios \red{1, 2, 4, and 5} have 90 virtual agents being simulated, with scenario 3 having 40 agents in total.

\red{Tables~\ref{table:average_ref_final} and~\ref{table:average_prime_eval} \red{present} the results obtained by our simulations. Table~\ref{table:average_ref_final} contains data regarding the Reference Simulation, which considers only one agent, and the Final Simulation, which considers all agents in the configuration, described in Equations~\ref{eq:averageTime} through~\ref{eq:averageDistance}. Table~\ref{table:average_prime_eval} contains data regarding the prime values, described in Equations~\ref{eq:normalTotalTime} through~\ref{eq:normalDistanceWalked}, and both evaluation metrics.}
S1-A refers to Scenario 1, Configuration A, and similarly to other scenarios and configurations.
The tables contain all five metrics required to calculate our proposed evaluation metric $\phi$ (column 6 in Table~\ref{table:average_prime_eval}). For convenience, the metric $\xi$, presented by Cassol et al.~\cite{cassol2017evaluating}, is also presented in the same table (column 7) to be used in comparisons (discussed in Section~\ref{sec:metricComparison}). 
\red{Table~\ref{table:average_prime_eval} also highlights in bold the best configurations for each scenario according to each evaluation metric.} 

Simulation Scenario 1 explores the movement of agents in different directions in an open space, with all agents positioned in the top-left corner and the goal positioned at different corners of the environment for each configuration. Our expectations were that S1-B and S1-C would have similar quantitative results and evaluation $\phi$, as agents would have to travel similar distances (although in different directions) in a straightforward path. We also expected that S1-A would have a higher value of $\phi$ (i.e., a worse evaluation) as the agents would have to travel a longer distance, increasing the values of $\bar{w}$, $\bar{t}$ and $t_g$. 
Indeed, as Table~\ref{table:average_prime_eval} presents, the simulation results met our expectations, with S1-B having the lowest value of \red{$\phi = 1.23$}, i.e., considered the best evacuation between the configurations, S1-C as the next best configuration with \red{$\phi = 1.24$}, and S1-A having the highest values of $t_g$, $\bar{t}$, and $\bar{w}$ (Table~\ref{table:average_ref_final}), and the worst evaluation with \red{$\phi = 1.32$}.

Simulation Scenario 2 was very similar to Scenario 1, where agents are initially placed in the same position in an open space. Agents move from the top-left corner towards the bottom-right, with the goal positioned at different distances in each configuration. Our expectations were that configurations with agents closer to their goal would have a better evaluation due to expected lower values of $t_g$, $\bar{t}$, and $\bar{w}$. This means that S2-A is expected to have the lowest value of $\phi$, followed by S2-C and then by S2-B. Analysing Table~\ref{table:average_prime_eval}, one can note that our expectations were achieved.

Simulation Scenarios \red{3, 4, and 5} represent similar conditions: the initial position of agents and the goals do not change in the three configurations; however, we changed the obstacles (red boxes). Regarding S3, we expected that S3-C should be the best one as a function of the simplicity of the environment, followed by S3-A. These expectations agree with $\phi$ values shown in Table~\ref{table:average_prime_eval}. With respect to S4, we expected that S4-B should present a lower value of $\phi$ due to fewer space restrictions imposed by the obstacles on the agents' motion. Again, our expectation was confirmed if we consider the $\phi$ values in Table~\ref{table:average_prime_eval}. \red{Regarding S5, our expectations were that S5-A would have the best evaluation, as agents could move directly toward their goals after passing the initial obstacles, and S5-B would have the worst evaluation due to the high amount of obstacles in the agents' paths. Once more, these expectations agree with the values presented in Table ~\ref{table:average_prime_eval}.}
In the next section, we compare the $\phi$ values obtained with our metric, and the Cassol et al.~\cite{cassol2017evaluating} $\xi$ metric.

\begin{table*}[!htb]
\scriptsize
\centering
\caption{\red{The table presents the averages and standard deviations of all metrics in each scenario configuration's reference and final simulations, considering 10 executions. The Simulation Scenarios are presented in Figures~\ref{fig:simulation_scenarios_part1} and~\ref{fig:simulation_scenarios_part2}. Columns present the metrics: the simulation time for the reference agent $t_{ar}$, the average speed of the reference agent $s_{ar}$, the average distance walked for the reference agent $w_{ar}$,
total simulation time $t_g$, average simulation time $\bar{t}$, average density $\bar{d}$, average speed $\bar{s}$, and average distance walked $\bar{w}$.}}
\begin{tabular}{@{}c|ccc|ccccc@{}}
\toprule
\multicolumn{1}{l}{} & \multicolumn{3}{c}{Reference Simulation}                                                                                                                                      & \multicolumn{5}{c}{Final Simulation}                                                                                                                                                                                                                                                               \\ \midrule
ID                   & $t_{ar}$                                                  & $s_{ar}$                                                  & $w_{ar}$                                                  & $t_g$                                                  & $\bar{t}$                                                  & $\bar{d}$                                                  & $\bar{s}$                                                  & $\bar{w}$                                                  \\ \midrule
S1-A                 & \begin{tabular}[c]{@{}c@{}}32.32\\ ($\pm$0.14)\end{tabular} & \begin{tabular}[c]{@{}c@{}}1.15\\ ($\pm$0.00)\end{tabular} & \begin{tabular}[c]{@{}c@{}}37.16\\ ($\pm$0.03)\end{tabular} & \begin{tabular}[c]{@{}c@{}}52.90\\ ($\pm$0.37)\end{tabular}   & \begin{tabular}[c]{@{}c@{}}40.54\\ ($\pm$0.21)\end{tabular}  & \begin{tabular}[c]{@{}c@{}}1.10\\ ($\pm$0.01)\end{tabular} & \begin{tabular}[c]{@{}c@{}}0.89\\ ($\pm$0.00)\end{tabular} & \begin{tabular}[c]{@{}c@{}}35.31\\ ($\pm$0.02)\end{tabular} \\
S1-B                 & \begin{tabular}[c]{@{}c@{}}23.01\\ ($\pm$0.12)\end{tabular} & \begin{tabular}[c]{@{}c@{}}1.15\\ ($\pm$0.01)\end{tabular} & \begin{tabular}[c]{@{}c@{}}26.52\\ ($\pm$0.03)\end{tabular} & \begin{tabular}[c]{@{}c@{}}43.93\\ ($\pm$0.66)\end{tabular}   & \begin{tabular}[c]{@{}c@{}}31.23\\ ($\pm$0.33)\end{tabular}  & \begin{tabular}[c]{@{}c@{}}1.14\\ ($\pm$0.00)\end{tabular} & \begin{tabular}[c]{@{}c@{}}0.84\\ ($\pm$0.01)\end{tabular} & \begin{tabular}[c]{@{}c@{}}25.13\\ ($\pm$0.03)\end{tabular} \\
S1-C                 & \begin{tabular}[c]{@{}c@{}}22.95\\ ($\pm$0.13)\end{tabular} & \begin{tabular}[c]{@{}c@{}}1.15\\ ($\pm$0.01)\end{tabular} & \begin{tabular}[c]{@{}c@{}}26.38\\ ($\pm$0.04)\end{tabular} & \begin{tabular}[c]{@{}c@{}}45.01\\ ($\pm$0.37)\end{tabular}   & \begin{tabular}[c]{@{}c@{}}31.81\\ ($\pm$0.21)\end{tabular}  & \begin{tabular}[c]{@{}c@{}}1.15\\ ($\pm$0.01)\end{tabular} & \begin{tabular}[c]{@{}c@{}}0.82\\ ($\pm$0.00)\end{tabular} & \begin{tabular}[c]{@{}c@{}}25.14\\ ($\pm$0.02)\end{tabular} \\ \midrule
S2-A                 & \begin{tabular}[c]{@{}c@{}}12.44\\ ($\pm$0.11)\end{tabular} & \begin{tabular}[c]{@{}c@{}}1.15\\ ($\pm$0.01)\end{tabular} & \begin{tabular}[c]{@{}c@{}}14.28\\ ($\pm$0.03)\end{tabular} & \begin{tabular}[c]{@{}c@{}}29.68\\ ($\pm$0.40)\end{tabular}   & \begin{tabular}[c]{@{}c@{}}18.75\\ ($\pm$0.25)\end{tabular}  & \begin{tabular}[c]{@{}c@{}}1.25\\ ($\pm$0.01)\end{tabular} & \begin{tabular}[c]{@{}c@{}}0.71\\ ($\pm$0.01)\end{tabular} & \begin{tabular}[c]{@{}c@{}}12.41\\ ($\pm$0.03)\end{tabular} \\
S2-B                 & \begin{tabular}[c]{@{}c@{}}32.32\\ ($\pm$0.14)\end{tabular} & \begin{tabular}[c]{@{}c@{}}1.15\\ ($\pm$0.00)\end{tabular} & \begin{tabular}[c]{@{}c@{}}37.16\\ ($\pm$0.03)\end{tabular} & \begin{tabular}[c]{@{}c@{}}52.95\\ ($\pm$0.33)\end{tabular}   & \begin{tabular}[c]{@{}c@{}}40.53\\ ($\pm$0.17)\end{tabular}  & \begin{tabular}[c]{@{}c@{}}1.10\\ ($\pm$0.00)\end{tabular} & \begin{tabular}[c]{@{}c@{}}0.89\\ ($\pm$0.00)\end{tabular} & \begin{tabular}[c]{@{}c@{}}35.31\\ ($\pm$0.02)\end{tabular} \\
S2-C                 & \begin{tabular}[c]{@{}c@{}}17.40\\ ($\pm$0.11)\end{tabular} & \begin{tabular}[c]{@{}c@{}}1.15\\ ($\pm$0.01)\end{tabular} & \begin{tabular}[c]{@{}c@{}}20.02\\ ($\pm$0.02)\end{tabular} & \begin{tabular}[c]{@{}c@{}}36.12\\ ($\pm$0.33)\end{tabular}   & \begin{tabular}[c]{@{}c@{}}24.60\\ ($\pm$0.22)\end{tabular}  & \begin{tabular}[c]{@{}c@{}}1.18\\ ($\pm$0.01)\end{tabular} & \begin{tabular}[c]{@{}c@{}}0.78\\ ($\pm$0.00)\end{tabular} & \begin{tabular}[c]{@{}c@{}}18.16\\ ($\pm$0.03)\end{tabular} \\ \midrule
S3-A                 & \begin{tabular}[c]{@{}c@{}}66.32\\ ($\pm$1.13)\end{tabular} & \begin{tabular}[c]{@{}c@{}}1.09\\ ($\pm$0.02)\end{tabular} & \begin{tabular}[c]{@{}c@{}}72.20\\ ($\pm$0.16)\end{tabular} & \begin{tabular}[c]{@{}c@{}}113.90\\ ($\pm$1.58)\end{tabular}  & \begin{tabular}[c]{@{}c@{}}89.29\\ ($\pm$0.62)\end{tabular}  & \begin{tabular}[c]{@{}c@{}}1.09\\ ($\pm$0.00)\end{tabular} & \begin{tabular}[c]{@{}c@{}}0.86\\ ($\pm$0.01)\end{tabular} & \begin{tabular}[c]{@{}c@{}}74.83\\ ($\pm$0.11)\end{tabular} \\
S3-B                 & \begin{tabular}[c]{@{}c@{}}71.06\\ ($\pm$0.72)\end{tabular} & \begin{tabular}[c]{@{}c@{}}1.10\\ ($\pm$0.01)\end{tabular} & \begin{tabular}[c]{@{}c@{}}78.29\\ ($\pm$0.19)\end{tabular} & \begin{tabular}[c]{@{}c@{}}132.91\\ ($\pm$12.18)\end{tabular} & \begin{tabular}[c]{@{}c@{}}99.09\\ ($\pm$0.76)\end{tabular}  & \begin{tabular}[c]{@{}c@{}}1.12\\ ($\pm$0.01)\end{tabular} & \begin{tabular}[c]{@{}c@{}}0.84\\ ($\pm$0.00)\end{tabular} & \begin{tabular}[c]{@{}c@{}}80.15\\ ($\pm$0.25)\end{tabular} \\
S3-C                 & \begin{tabular}[c]{@{}c@{}}39.42\\ ($\pm$0.17)\end{tabular} & \begin{tabular}[c]{@{}c@{}}1.14\\ ($\pm$0.00)\end{tabular} & \begin{tabular}[c]{@{}c@{}}45.08\\ ($\pm$0.09)\end{tabular} & \begin{tabular}[c]{@{}c@{}}62.48\\ ($\pm$0.85)\end{tabular}   & \begin{tabular}[c]{@{}c@{}}49.70\\ ($\pm$0.34)\end{tabular}  & \begin{tabular}[c]{@{}c@{}}1.07\\ ($\pm$0.00)\end{tabular} & \begin{tabular}[c]{@{}c@{}}0.92\\ ($\pm$0.01)\end{tabular} & \begin{tabular}[c]{@{}c@{}}44.94\\ ($\pm$0.05)\end{tabular} \\ \midrule
S4-A                 & \begin{tabular}[c]{@{}c@{}}56.30\\ ($\pm$0.68)\end{tabular} & \begin{tabular}[c]{@{}c@{}}1.12\\ ($\pm$0.01)\end{tabular} & \begin{tabular}[c]{@{}c@{}}63.24\\ ($\pm$0.16)\end{tabular} & \begin{tabular}[c]{@{}c@{}}147.00\\ ($\pm$15.29)\end{tabular} & \begin{tabular}[c]{@{}c@{}}100.89\\ ($\pm$0.61)\end{tabular} & \begin{tabular}[c]{@{}c@{}}1.28\\ ($\pm$0.02)\end{tabular} & \begin{tabular}[c]{@{}c@{}}0.68\\ ($\pm$0.01)\end{tabular} & \begin{tabular}[c]{@{}c@{}}63.80\\ ($\pm$0.40)\end{tabular} \\
S4-B                 & \begin{tabular}[c]{@{}c@{}}47.56\\ ($\pm$0.61)\end{tabular} & \begin{tabular}[c]{@{}c@{}}1.10\\ ($\pm$0.01)\end{tabular} & \begin{tabular}[c]{@{}c@{}}52.32\\ ($\pm$0.14)\end{tabular} & \begin{tabular}[c]{@{}c@{}}137.70\\ ($\pm$6.78)\end{tabular}  & \begin{tabular}[c]{@{}c@{}}92.65\\ ($\pm$0.81)\end{tabular}  & \begin{tabular}[c]{@{}c@{}}1.32\\ ($\pm$0.02)\end{tabular} & \begin{tabular}[c]{@{}c@{}}0.62\\ ($\pm$0.00)\end{tabular} & \begin{tabular}[c]{@{}c@{}}53.60\\ ($\pm$0.40)\end{tabular} \\
S4-C                 & \begin{tabular}[c]{@{}c@{}}55.61\\ ($\pm$0.41)\end{tabular} & \begin{tabular}[c]{@{}c@{}}1.14\\ ($\pm$0.01)\end{tabular} & \begin{tabular}[c]{@{}c@{}}63.26\\ ($\pm$0.08)\end{tabular} & \begin{tabular}[c]{@{}c@{}}143.62\\ ($\pm$9.44)\end{tabular}  & \begin{tabular}[c]{@{}c@{}}99.50\\ ($\pm$1.24)\end{tabular}  & \begin{tabular}[c]{@{}c@{}}1.29\\ ($\pm$0.01)\end{tabular} & \begin{tabular}[c]{@{}c@{}}0.69\\ ($\pm$0.01)\end{tabular} & \begin{tabular}[c]{@{}c@{}}64.21\\ ($\pm$0.11)\end{tabular} \\ \midrule
S5-A                 & \begin{tabular}[c]{@{}c@{}}47.87\\ ($\pm$0.15)\end{tabular} & \begin{tabular}[c]{@{}c@{}}1.15\\ ($\pm$0.00)\end{tabular} & \begin{tabular}[c]{@{}c@{}}55.10\\ ($\pm$0.05)\end{tabular} & \begin{tabular}[c]{@{}c@{}}78.15\\ ($\pm$0.57)\end{tabular}   & \begin{tabular}[c]{@{}c@{}}61.28\\ ($\pm$0.31)\end{tabular}  & \begin{tabular}[c]{@{}c@{}}1.10\\ ($\pm$0.00)\end{tabular} & \begin{tabular}[c]{@{}c@{}}0.89\\ ($\pm$0.00)\end{tabular} & \begin{tabular}[c]{@{}c@{}}53.45\\ ($\pm$0.04)\end{tabular} \\
S5-B                 & \begin{tabular}[c]{@{}c@{}}58.80\\ ($\pm$0.32)\end{tabular} & \begin{tabular}[c]{@{}c@{}}1.14\\ ($\pm$0.00)\end{tabular} & \begin{tabular}[c]{@{}c@{}}66.93\\ ($\pm$0.10)\end{tabular} & \begin{tabular}[c]{@{}c@{}}142.34\\ ($\pm$0.78)\end{tabular}  & \begin{tabular}[c]{@{}c@{}}102.51\\ ($\pm$0.77)\end{tabular} & \begin{tabular}[c]{@{}c@{}}1.21\\ ($\pm$0.01)\end{tabular} & \begin{tabular}[c]{@{}c@{}}0.70\\ ($\pm$0.00)\end{tabular} & \begin{tabular}[c]{@{}c@{}}68.61\\ ($\pm$0.11)\end{tabular} \\
S5-C                 & \begin{tabular}[c]{@{}c@{}}48.45\\ ($\pm$0.13)\end{tabular} & \begin{tabular}[c]{@{}c@{}}1.15\\ ($\pm$0.00)\end{tabular} & \begin{tabular}[c]{@{}c@{}}55.87\\ ($\pm$0.08)\end{tabular} & \begin{tabular}[c]{@{}c@{}}90.54\\ ($\pm$2.47)\end{tabular}   & \begin{tabular}[c]{@{}c@{}}64.51\\ ($\pm$0.28)\end{tabular}  & \begin{tabular}[c]{@{}c@{}}1.05\\ ($\pm$0.01)\end{tabular} & \begin{tabular}[c]{@{}c@{}}0.87\\ ($\pm$0.01)\end{tabular} & \begin{tabular}[c]{@{}c@{}}54.06\\ ($\pm$0.48)\end{tabular} \\ \bottomrule
\end{tabular}
\label{table:average_ref_final}
\end{table*}

\begin{table*}[htb]
\scriptsize
\centering
\caption{\red{The table presents the averages and standard deviations of each scenario configuration's prime and evaluation metrics, considering 10 executions. The Simulation Scenarios are presented in Figures~\ref{fig:simulation_scenarios_part1} and~\ref{fig:simulation_scenarios_part2}. Columns present the metrics: the prime simulation time $t^{'}_g$, the prime average simulation time $t^{'}$, the prime average speed $s^{'}$, the prime average distance walked $s^{'}$,  our proposed evaluation metric $\phi$, and the metric proposed by Cassol et al.~\cite{cassol2017evaluating} $\xi$. In bold, lower values of $\phi$ and $\xi$ represent the best evacuation scenarios according to each metric.}}
\begin{tabular}{@{}c|cccc|cc@{}}
\toprule
\multicolumn{1}{l}{} & \multicolumn{4}{c}{Prime Variables}                                               & \multicolumn{2}{c}{Evaluation Metrics} \\ \midrule
ID                   & $t^{'}_g$ & $\bar{t^{'}}$ & $\bar{s^{'}}$ & $\bar{w^{'}}$ & $\phi$ (Ours)              & $\xi$~\cite{cassol2017evaluating}      \\ \midrule
S1-A                 & 1.63 ($\pm$0.01)       & 1.25 ($\pm$0.01)            & 3.61 ($\pm$0.02)  & 0.83 ($\pm$0.00)            & 1.32 ($\pm$0.00)       & \textbf{1.54 ($\pm$0.01)}      \\
S1-B                 & 1.92 ($\pm$0.03)       & 1.36 ($\pm$0.01)            & 3.98  ($\pm$0.05) & 0.59 ($\pm$0.00)            & \textbf{1.23 ($\pm$0.01)}       & 1.68 ($\pm$0.01)      \\
S1-C                 & 1.98 ($\pm$0.02)       & 1.40 ($\pm$0.01)            & 4.07 ($\pm$0.03)  & 0.59 ($\pm$0.00)            & 1.24 ($\pm$0.00)       & 1.71 ($\pm$0.01)      \\ \midrule
S2-A                 & 2.37 ($\pm$0.03)       & 1.49 ($\pm$0.02)            & 4.93 ($\pm$0.08)  & 0.29 ($\pm$0.00)            & \textbf{0.91 ($\pm$0.00)}       & 1.91 ($\pm$0.02)      \\
S2-B                 & 1.64 ($\pm$0.01)       & 1.25 ($\pm$0.01)            & 3.61 ($\pm$0.02)  & 0.83 ($\pm$0.00)            & 1.32 ($\pm$0.00)       & \textbf{1.54 ($\pm$0.00)}      \\
S2-C                 & 2.07 ($\pm$0.02)       & 1.41 ($\pm$0.01)            & 4.36 ($\pm$0.04)  & 0.43 ($\pm$0.00)            & 1.09 ($\pm$0.00)       & 1.76 ($\pm$0.01)      \\ \midrule
S3-A                 & 1.72 ($\pm$0.03)       & 1.35 ($\pm$0.02)            & 3.56 ($\pm$0.05)  & 1.76 ($\pm$0.00)            & 1.62 ($\pm$0.01)       & 1.59 ($\pm$0.01)      \\
S3-B                 & 1.88 ($\pm$0.17)       & 1.40 ($\pm$0.01)            & 3.74 ($\pm$0.03)  & 1.89 ($\pm$0.01)            & 1.70 ($\pm$0.02)       & 1.66 ($\pm$0.03)      \\
S3-C                 & 1.59 ($\pm$0.02)       & 1.26 ($\pm$0.01)            & 3.47 ($\pm$0.03)  & 1.06 ($\pm$0.00)            & \textbf{1.39 ($\pm$0.01)}       & \textbf{1.51 ($\pm$0.01) }     \\ \midrule
S4-A                 & 2.65 ($\pm$0.28)       & 1.82 ($\pm$0.01)            & 5.40 ($\pm$0.07)  & 1.50 ($\pm$0.01)            & 1.95 ($\pm$0.02)       & 2.11 ($\pm$0.03)      \\
S4-B                 & 2.87 ($\pm$0.14)       & 1.93 ($\pm$0.02)            & 5.70 ($\pm$0.04)  & 1.26 ($\pm$0.01)            & \textbf{1.93 ($\pm$0.01)}       & 2.22 ($\pm$0.01)      \\
S4-C                 & 2.57 ($\pm$0.17)       & 1.78 ($\pm$0.02)            & 5.19 ($\pm$0.08)  & 1.51 ($\pm$0.00)            & 1.94 ($\pm$0.02)       & \textbf{2.08 ($\pm$0.03)}      \\ \midrule
S5-A                 & 1.63 ($\pm$0.01)       & 1.28 ($\pm$0.01)            & 3.64 ($\pm$0.02)  & 0.86 ($\pm$0.00)            & \textbf{1.34 ($\pm$0.00)}       & \textbf{1.55 ($\pm$0.01)}      \\
S5-B                 & 2.42 ($\pm$0.01)       & 1.74 ($\pm$0.01)            & 5.03 ($\pm$0.05)  & 1.11 ($\pm$0.00)            & 1.72 ($\pm$0.01)       & 1.99 ($\pm$0.01)      \\
S5-C                 & 1.87 ($\pm$0.05)       & 1.33 ($\pm$0.01)            & 3.80 ($\pm$0.05)  & 0.87 ($\pm$0.01)            & 1.37 ($\pm$0.00)       & 1.60 ($\pm$0.01)      \\ \bottomrule
\end{tabular}
\label{table:average_prime_eval}
\end{table*}

\subsection{Comparison $\phi$ with $\xi$ evaluation metrics}
\label{sec:metricComparison}

In this section, we compare our proposed evaluation metric ($\phi$) with the one presented in the literature by Cassol et al.~\cite{cassol2017evaluating} ($\xi$).
Firstly, it is important to highlight that although both evaluation metrics $\phi$ and $\xi$ present scalar values, their values should not be directly compared. For example, S1-A has the values of \red{$\phi = 1.32$} and \red{$\xi = 1.54$}. This does not indicate that our metric $\phi$ considers S1-A a better configuration when compared to $\xi$. Instead, values should be compared between different configurations. In the case of Simulation Scenario 1, $\phi$ indicates that configuration S1-B is the better option due to having a lower value than other configurations, whereas $\xi$ indicates that configuration S1-A is the better option.

As seen in Table~\ref{table:average_prime_eval}, in the last two columns, the best configuration for each evaluation metric are only in accordance with S3-C \red{and S5-A}.
Analyzing scenarios 1 and 2, we can observe that metric $\xi$ proposed by Cassol et al.~\cite{cassol2017evaluating} selected as the best ones S1-A and S2-B, which are the configurations where people are located far from the goal when compared with other configurations from the same scenario. Although this fact also negatively impacts the simulation time factors, it improves densities and speeds to compute the $\xi$ metric. On the other hand, in our metric $\phi$, while speeds and densities are positively impacted as well (see columns 7,8 in Table~\ref{table:average_ref_final}), traveled distance counts negatively for the metric, so S1-A and S2-B are the worst evaluated scenarios, according to $\phi$.

Regarding S3, it is easy to see that the best crowd scenario is S3-C,  
due to the reduced number of obstacles and increased space available for agents. \red{It happens because having lower average simulation times, density, and distance walked, while also having higher average speed, positively impacts $\phi$ and $\xi$ metrics.}
S4 presents a similar case such as S1. While the $\xi$ metric selects the higher speed and lower density metrics even if time factors are higher (S4-A), the $\phi$ metric selects the smaller time factors and less traveled distance.
\red{S5 presents the only case where $\xi$ and $\phi$ agree on the best, mid, and worst configurations. Although S5-A presents a higher average density ($\bar{d}$) than S5-C, all other values positively impact both evaluation metrics.}

\subsection{Experiment with Subjects}
\label{sec:focusGroup}


In addition to the comparison with the literature, which presents a quantitative assessment of our metric  $\phi$, we performed a qualitative evaluation with a group of \red{experts in crowd scenarios (CS)}. We invited three experts who work with crowd behaviors. Indeed, experts 1 and 2 work with crowd simulation, while expert 3 works with computer vision applied for crowds. \red{CS experts} were presented with an online questionnaire, starting with a description of our project, the estimated duration, and the authors' contact information. Participants were asked to grant access to their answers and personal information following ethical guidelines\footnote{Project ``Estudos e Avaliações da Percepção Humana em Personagens e Multidões Virtuais'', number 46571721.6.0000.5336, approved by the Ethics Committee of Pontifical Catholic University of Rio Grande do Sul.}. This includes information such as age, gender, and educational level. All experts are male, aged between 25 and 35, and work in Computer Science.

Given their agreement, participants were presented with \red{five} tasks, one for each of the simulation scenarios presented in Figures~\ref{fig:simulation_scenarios_part1} and~\ref{fig:simulation_scenarios_part2}. Each task required participants to rank the three different crowd configurations \red{of each scenario}, based on their experience, from best to worst evacuation plan. The task contained an image of each configuration, the same as in Figures~\ref{fig:simulation_scenarios_part1} and~\ref{fig:simulation_scenarios_part2}, but without the Occupancy Map and the Agents' Trajectories, i.e., only the first lines of each scenario. 
The task also contained the information of columns 5-9 of Table~\ref{table:average_ref_final}, presenting the Final Simulation metrics for each configuration without including the evaluation metrics $\phi$ and $\xi$.

Table~\ref{table:experts_results} presents the answers of the three \red{CS experts} (i.e., Expert 1, Expert 2, and Expert 3) for the \red{five} scenarios proposed (i.e., S1, S2, S3, \red{S4, and S5}). For each scenario, three configurations (i.e., A, B, and C) were presented to the experts and ranked from best to worst evacuation scenario (i.e., Best, Mid, and Worst). 
The second column of Table~\ref{table:experts_results} shows the ranking as calculated for our metric ($\phi$) and Cassol's metric ($\xi$). It is possible to note that the majority of the \red{CS experts}' answers agree with our metric $\phi$, with the exceptions being Expert 3 in Scenarios S1 and S2, which agrees with the $\xi$ metric\red{, Expert 1 disagreeing with both metrics for the best and mid configurations in S5, and all experts disagreeing with our metric $\phi$ for the mid and worst configurations of S4}

\begin{table}[!htb]
\centering
\caption{The table presents the evaluation of scenarios. \red{S1, S2, S3, S4 and S5 state for the five tested scenarios}. The evaluations Best/Mid/Worst present the rank of the three configurations according to our metric ($\phi$ in the $2^{nd}$ column), $\xi$ metric in $3^{rd}$ column, and the three \red{crowd scenario (CS) experts} in the other columns. Experts' answers in bold agree with the $\phi$ metric, while underlined are the scenarios that agree with the $\xi$ metric.
}
\begin{tabular}{@{}cc|cc|ccc@{}}
\toprule
\multicolumn{2}{c|}{Scenarios} & $\phi$ & $\xi$ & Expert 1 
& Expert 2 
& Expert 3 
\\ \midrule
\multirow{3}{*}{S1}   & Best   & B      & A     & \textbf{B}     & \textbf{B}     & \underline{A}                               \\
                      & Mid    & C      & B     & \textbf{C}     & \textbf{C}    & \underline{B}      \\
                      & Worst  & A      & C     & \textbf{A}     & \textbf{A}     & \underline{C}                                \\ \midrule
\multirow{3}{*}{S2}   & Best   & A      & B     & \textbf{A}     & \textbf{A}     & \underline{B}                               \\
                      & Mid    & C      & C     & \textbf{C}     & \textbf{C}     & \underline{C}                               \\
                      & Worst  & B      & A     & \textbf{B}     & \textbf{B}     & \underline{A}                               \\ \midrule
\multirow{3}{*}{S3}   & Best   & C      & C     & \underline{\textbf{C}}    & \underline{\textbf{C}}     & \underline{\textbf{C}}      \\
                      & Mid    & A      & A     & \underline{\textbf{A}}     & \underline{\textbf{A}}     & \underline{\textbf{A}}                               \\
                      & Worst  & B      & B     & \underline{\textbf{B}}     & \underline{\textbf{B}}     & \underline{\textbf{B}}                               \\ \midrule
\multirow{3}{*}{S4}   & Best   & B      & C     & \textbf{B}     & \textbf{B}     & \textbf{B}                               \\
                      & Mid    & C      & A     & \underline{A}     & \underline{A}      & \underline{A}                                \\
                      & Worst  & A      & B     & C     & C     & C                              \\ \midrule
                      
\multirow{3}{*}{\red{S5}}   & \red{Best}   & A      & A     & C    & \underline{\textbf{A}}     & \underline{\textbf{A}}                               \\
                      & \red{Mid}    & C      & C     & A     & \underline{\textbf{C}}    & \underline{\textbf{C}}                               \\
                      & \red{Worst}  & B      & B     & \underline{\textbf{B}}     & \underline{\textbf{B}}     & \underline{\textbf{B}}                               \\ \bottomrule
\end{tabular}
\label{table:experts_results}
\end{table}

While more investigation is needed in the future to understand the difference between \red{CS experts}' opinions, we hypothesize here that the fact that experts 1 and 2 work with crowd simulation, they intuitively give relevance to distance features, while the computer vision expert considered more the density and speed features. 
\red{Additionally, each configuration's visual information could have impacted the ranking of experts. Expert 1 ranked S5-A, which contains two obstacles that heavily restrict the movement of agents at the beginning of the simulation, as worse 
than S5-C, even though S5-A has four metrics with better values when compared to S5-C, except the average density.}
\red{In regards to the experts disagreeing with our metric $\phi$ for the mid and worst configurations of S4, we believe that the high standard deviations for the total simulation time $t_g$ and average simulation time $\bar{t}$ of S4-A and S4-C (columns 5 and 6 in Table~\ref{table:average_ref_final}) could have caused the difference in ranking when considering multiple executions of the same simulation scenario.}

\section{Final Considerations}
\label{sec:finalConsiderations}

This paper presented WebCrowds, an authoring tool for crowd simulation that anyone can use to build environments and simulate the movement of agents. In addition, WebCrowds provides a way to evaluate the crowd scenarios and compare configurations of the same environment to determine the best configuration for the crowd. This work is relevant because selecting the best evacuation plan is a problem in safety engineering~\cite{cassol2017evaluating,DBLP:journals/jvca/UsmanHFK21} Therefore, a straightforward tool that can be used by anyone, as shown in~\cite{silva2022webcrowds}, is interesting to provide space studies.

To evaluate and compare the crowd configurations, we propose a new evaluation metric $\phi$ that congregates together crowd features, such as speeds, time factors, and densities, in particular considering the traveled distance by agents. We compared $\phi$ with a crowd evaluation metric proposed in the literature ($\xi$) and also with a focus group of \red{CS experts}. Our method 
showed more agreement with the subjective opinion of the experts.



WebCrowds has some limitations. Firstly, in a building with more than one floor, configuration scenarios should be modeled separately for each one. Also, modeling a complex and big environment should be challenging for the user. We want to work on these usability topics in a future version. In addition, we want to work on a mobile application where people can model their environment (store, restaurant, nightclub), and the public can have access to understand the ways of evacuating the environment. 
\red{Another limitation concerns the experts used in our experiments, two from crowd simulation and one from computer vision. It would be interesting to have more experts from different fields, such as crowd evacuation, police, and firefighters, who could be interested in such a tool. Moreover, besides evaluating the best, mid, and worst configurations, it could be interesting to ask the participants what they are considering to reach their conclusions. It can also be included in a future work.}


\section*{Acknowledgments}
This study was partly financed by the Coordenação de Aperfeiçoamento de Pessoal de Nivel Superior – Brasil (CAPES) – Finance Code 001, and
by the Conselho Nacional de Desenvolvimento Científico e Tecnológico - Brasil (CNPq).
We thank the experts who participated in our experiments.
Finally, we also thank the High-Performance Computing Laboratory of the Pontifical Catholic University of Rio Grande do Sul (LAD-IDEIA/PUCRS) for providing support and technological resources for this project.



 \bibliographystyle{elsarticle-num} 
 \bibliography{cas-refs}





\end{document}